\pdfoutput=1
\documentclass[a4paper,11pt]{article}

\usepackage{fullpage}
\usepackage{dsfont,xspace}
\usepackage{amsmath,amsthm,amssymb}
\usepackage{graphicx,subfigure,caption}
\usepackage{graphicx}
\usepackage{caption}
\usepackage{color}
\usepackage{todonotes}

\newcommand{\ie}{{\emph{i.e.\/}}}
\newcommand{\etal}{\emph{et al.}}
\providecommand{\keywords}[1]{\newline Keywords: #1}
\providecommand{\sep}{\hspace{-3pt};\xspace}

\newcommand{\halmos}{\par\hfill $\Box$\vspace{6pt}}

\providecommand{\proof}{\noindent{\emph{Proof.}}}
\providecommand{\proofof}[1]{\vspace{12pt}\noindent{\emph{Proof of #1.}}}

\newcommand{\LieG}[2]{\ensuremath{\mathds{#1}(#2)}}
\newcommand{\SU}[1]{\LieG{SU}{#1}}

\newtheorem{definition}{Definition}

\DeclareMathOperator{\tr}{tr}

\newcommand{\C}{\ensuremath{\mathds{C}}}
\newcommand{\boldone}{\ensuremath{\mathds{1}}}
\newcommand{\Cplx}{\ensuremath{\C}}

\newcommand{\ket}[1]{\ensuremath{|#1\rangle}}
\newcommand{\bra}[1]{\ensuremath{\langle#1|}}
\newcommand{\ketbra}[2]{\ensuremath{\ket{#1}\bra{#2}}}
\newcommand{\braket}[2]{\ensuremath{\langle{#1}|{#2}\rangle}}
\newcommand{\Id}{\mathds{1}}
\newcommand{\floor}[1]{\ensuremath{\lfloor #1 \rfloor}}
\newcommand{\complexity}[1]{\ensuremath{\mathbf{#1}}}

\newtheorem{fact}{Fact}

\newtheorem{proposition}{Proposition}

\newcounter{strategy}
\newenvironment{strategy}
{\refstepcounter{strategy}\par\medskip\noindent\textbf{Strategy \arabic{strategy}.}}
{\par\medskip}

\begin{document}

\title{Quantum network exploration with a faulty sense of direction}

\author{Jaros{\l}aw Adam Miszczak\footnote{Electronic address: \texttt{miszczak@iitis.pl}} \quad Przemys{\l}aw Sadowski\footnote{Electronic address: \texttt{psadowski@iitis.pl}}
\\Institute of Theoretical and Applied Informatics,\\ Polish Academy of Sciences,
\\ Ba{\l}tycka 5, 44-100 Gliwice, Poland}

\date{25/03/2014 (v. 0.43)}

\maketitle

\begin{abstract}
We develop a model which can be used to analyse the scenario of exploring
quantum network with a distracted sense of direction. Using this model we
analyse the behaviour of quantum mobile agents operating with non-adaptive and
adaptive strategies which can be employed in this scenario. We introduce the
notion of node visiting suitable for analysing quantum superpositions of states
by distinguishing between visiting and attaining a position. We show that
without a proper model of adaptiveness, it is not possible for the party
representing the distraction in the sense of direction, to obtain the results
analogous to the classical case. Moreover, with additional control resources the
total number of attained positions is maintained if the number of visited
positions is strictly limited.
\\
\end{abstract}


\section{Introduction}
Recent progress in quantum communication technology has confirmed that the
biggest challenge in using quantum methods of communication is to provide
scalable methods for building large-scale quantum
networks~\cite{gisin11technology, clausen11storage, pembertonross11perfect}. The
problems arising in this area are related to physical realizations of such
networks, as well as to designing new protocols that exploit new possibilities
offered by the principles of quantum mechanics in long-distance communication.

One of the interesting problems arising in the area of quantum internetworking
protocols is the development of methods which can be used to detect errors that
occur in large-scale quantum networks. A natural approach for developing such
methods is to construct them on the basis of the methods developed for classical
networks~\cite{vanmeter11recursive,munro11designing}.

The main contribution of this paper is the development of a method for exploring
quantum networks by mobile agents which operate on the basis of information
stored in quantum registers. We construct a model based on a quantum walk on
cycle which can be applied to analyse the scenario of exploring quantum networks
with a faulty sense of direction. One should note that the presented model
allows studying the situations where all nodes in the network are connected. The
reason for this is that a move can result in the shift of the token from the
current position to any other position in the network. Thus we do not restrict
ourselves to a cycle topology.

This paper is organized as follows.
In the remaining part of this Section we provide a motivation for the considered
scenario and recall a classical scenario described by Magnus-Derek game.
In Section~\ref{sec:quantum-magnus-derek} we introduce a quantum the scenario of
quantum network exploration with a distracted sense of direction. 
In Section~\ref{sec:application-quantum} we analyse the behaviour of quantum
mobile agents operating with various classes of strategies and describe
non-adaptive and adaptive quantum strategies which can be employed by the
players.
Finally, in Section~\ref{sec:final} we summarize the presented work and provide
some concluding remarks.

\subsection{Motivation}\label{sec:motivation}
As quantum networks consist of a large number of independent
parties~\cite{kimble08quantum, chapuran09optical} it is crucial to understand
how the errors, that occur during the computation on nodes, influence their
behaviour. Such errors may arise, in the first place, due to the erroneous work
of particular nodes. Therefore it is important to develop the methods that allow
the exploration of quantum networks and the detection of malfunctioning nodes.

One of the methods used to tackle this problem in classical networks is the
application of mobile agents, \ie\ autonomous computer programs which move
between hosts in a network. This method has been studied extensively in the
context of intrusion detection \cite{bernardes00implementation,
jansen02intrusion}, but it is also used as a convincing programming paradigm in
other areas of software engineering~\cite{lange99seven}. 

On the other hand, recent results concerning the exploration of quantum graphs
suggest that by using the rules of quantum mechanics it is possible to solve
search problems~\cite{reitzner09searches} or rapidly detect errors in
graphs~\cite{feldman10structural}.

In this paper we aim to combine both methods mentioned above. We focus on a
model of mobile agents used to explore a quantum network. For the purpose of
modelling such agents we introduce and study the quantum version of the
Magnus-Derek game~\cite{nedev08magnus-derek}. This combinatorial game,
introduced in~\cite{nedev08magnus-derek}, provides a model for describing a
mobile agent acting in a communication network. 

\subsection{Preliminaries}\label{sec:preliminaries}
The Magnus-Derek game was introduced in \cite{nedev08magnus-derek} and analysed
further in~\cite{hurkens08revisited} and~\cite{chen11more}. The game is played
by two players: Derek (from \emph{direction} or \emph{distraction}) and Magnus
(from \emph{magnitude} or \emph{maximization}), who operate by moving a token on
a round table (cycle) with $n$ nodes $0,1,\ldots,n-1$. Initially the token is
placed in the position $0$. In each round (step) Magnus decides about the number
$0\leq m\leq \frac{n}{2}$ of positions for the token to move and Derek decides
about the direction: clockwise ($+$ or $0$) or counter-clockwise ($-$ or $1$). 

Magnus aims to maximize the number of nodes visited during the game, while Derek
aims to minimize this value. Derek represents a distraction in the sense of
direction. For example, a~sequence of moves $0\rightarrow 1 \rightarrow 2$
allowing Magnus to visit three nodes, can be changed to $0
\stackrel{+}{\rightarrow} 1 \stackrel{-}{\rightarrow} 0$ due to the influence of
Derek represented by the $+$ and $-$ signs. The possibility of providing biased
information about the direction prevents Magnus permanently from visiting some
nodes.

In the classical scenario one can introduce a function $f^\star(n)$ which, for a
given number of nodes $n$, gives the cardinality of the set of positions visited
by the token when both players play optimally~\cite{nedev08magnus-derek}. It can
be shown that this function is well defined and 
\begin{equation}
f^\star(n)=
\left\{\begin{array}{cl}
    n &\ \mathrm{for}\ n=2^k,\\
    \frac{(p-1)n}{p} &\ \mathrm{for}\ n=pm,
\end{array}\right.
\end{equation}
with $p$ being the smallest odd prime factor of $n$.

By $r(n)$ we denote the number of moves required to visit the optimal number of
nodes. In the case $n=2^k$, the number of moves is optimal and equals
$r(2^k)=2^k-1$. Hurkens \etal\ proved~\cite{hurkens08revisited} that if $n\geq3$
is a positive integer not equal to a power of $2$, then there exists a strategy
allowing Magnus to visit at least $f^\star(n)$ nodes using at most
$f^\star(n)\lceil\log_2(n-1)\rceil$ moves.

We distinguish two main types of regimes -- adaptive and non-adaptive. In the
adaptive regime, both players are able to choose their moves during the
execution of the game. In the non-adaptive regime, Magnus announces the sequence
of moves he aims to perform. In particular, if the game is executed in the
non-adaptive regime, Derek can calculate his sequence of moves before the game.
In the classical case the problem of finding the optimal strategy for Derek is
$\complexity{NP}$-hard~\cite{chen11more} and is equivalent to the partition
problem~\cite{papadimitriou}.

\section{Exploration of quantum networks}\label{sec:quantum-magnus-derek}
Let us now assume that the players operate by encoding their positions on a
cycle in an $n$-dimensional pure quantum states. Thus the position of the token
is encoded in a state $\ket{x}\in\Cplx^n$. At the $i$-th step of the game 
Magnus
decides to move $0\leq m_i\leq \frac{n}{2}$ and Derek decides to move in
direction $d_i\in\{0,1\}$.

One can easily express the classical game by applying the notation of quantum states.
The evolution of the system during the move described above is given by a
unitary matrix of the form
\begin{equation}
A^{(c)}_i= \sum_{k=0}^{n-1} \ketbra{k+(-1)^{d_i} m_i\ (\mathrm{mod}\ n)}{k},
\end{equation}
where $i=1,\dots, r(n)$. Clearly, as the above permutation operators express
only the classical subset of the possible moves, by using it one cannot expect
to gain with respect to the classical scenario. In particular, the operators
$A^{(c)}_i$ as introduced above do not allow the preparation of a move by using
the information encoded in a superposition.

In order to exploit the possibilities offered by quantum mechanics in the
Magnus-Derek scheme, we can use a quantum walk controlled by two registers. To
achieve this we need to offer the players a larger state space. We introduce a
quantum scheme by defining the following quantum version of the Magnus-Derek
game.
\begin{enumerate}
    \item The state of the system is described by a vector of the form
    \begin{equation}
        \ket{m}\ket{d}\ket{x}\in\Cplx^{\floor{n/2}}\otimes\Cplx^2\otimes\Cplx^n.
    \end{equation}
    
    \item The initial state of the system reads $\ket{\psi_0}=\ket{0\ldots 0}$.

    \item At each step the players can choose their strategy, possibly using
    unitary gates.
    \begin{enumerate}

    \item Magnus operates on his register with any unitary gate
    $M_i\in\SU{\floor{n/2}}$ resulting in a operation of the form
    $M_i\otimes\Id_2\otimes\Id_n$ performed on the full system.

    \item Derek operates on his register with any unitary gate
    $D_i\in\SU{2}$.
    If his actions are position-independent the operation performed on 
    the full system takes the form $\Id_{\floor{n/2}}\otimes D_i\otimes\Id_n$.
    However, in Section~\ref{sec:position_control} we also allow
    position-controlled actions, resulting in the operator of the form
    $\sum_x\Id_{\floor{n/2}}\otimes D_i^{(x)}\otimes\Pi_{x}$.
    \label{game:3b}
    \end{enumerate}

    \item The change of the token position, resulting from the players moves, is
    described by the shift operator
    \begin{equation}
    S=\sum_{m=0}^{\floor{n/2}}\sum_{k=0}^n \ketbra{m,0}{m,0}\otimes\ketbra{k+m}{k} +
    \sum_{m=0}^{\floor{n/2}}\sum_{k=0}^n\ketbra{m,1}{m,1}\otimes\ketbra{k-m}{k},
%
    \end{equation}
\end{enumerate}
where the addition and the subtraction is in the appropriate ring
$\mathds{Z}_n$.

The single move in the game defined according to the above description is given
by the position-independent operator 
\begin{equation}
A_i=S(M_i\otimes D_i\otimes \Id_n).
\end{equation}
Taking this into account the state of the system after the execution of $k$
moves reads
\begin{equation}
\ket{\psi_k} = A_k \ldots A_2 A_1 \ket{\psi_0},
\end{equation}
where each matrix $A_i,i=1,2,\ldots, k$ depends on the move of each party. The
distribution of the position on the cycle after $k$ moves is described by a
reduced density matrix
\begin{equation}
\rho_k = \tr_{M,D}(\ketbra{\psi_k}{\psi_k})=
        \tr_{\Cplx^{\floor{n/2}}\otimes\Cplx^{2}}(\ketbra{\psi_k}{\psi_k}),
\end{equation}
which represents the state of the token register after tracing-out the
subsystems used to process the strategies. Here $\tr_{M,D}(\cdot)$ represents
the operation of tracing-out the subsystems used by Magnus and Derek to encode
their strategies.

The key part of this procedure is how the players choose their strategies. The
selection of the method influences the efficiency of the exploration. Below we
study the possible methods and show how they influence the behaviour of the
quantum version of the Magnus-Derek game. 

Clearly, by using the unitary gates Magnus and Derek are able to prepare the
superpositions of base states. For this reason, one needs to provide the notion
of node visiting suitable for analysing quantum superpositions of states.
Therefore, we introduce the notion of \emph{visiting} and \emph{attaining} a
position.

\begin{definition}
We say that the position $x$ is visited in $t$ steps, if for some step $i\leq 
t$
the probability of measuring the position register in the state $\ket{x}$ is 1,
\ie\
\begin{equation}
\tr \ketbra{x}{x} (\tr_{M,D}\ketbra{\psi_i}{\psi_i}) =1.
\end{equation}
\end{definition}

In order to introduce the notion of attaining we use the concepts of measured
quantum walk~\cite{kempe05discrete} and concurrent hitting time.

\begin{definition}
A $\ket{x}$-measured quantum walk from a discrete-time quantum walk starting 
in a state $\ket{\psi_0}$ is a process defined by iteratively first measuring 
with the two projectors $\Pi_0 = \ket{x}\bra{x}$ and $\Pi_1 = \boldone - 
\Pi_0$. If 
$\Pi_0$ is measured the process is stopped, otherwise a step operator is 
applied and the iteration is continued.
\end{definition}

\begin{definition} 
A quantum random walk has a $(T, p)$ concurrent $(\ket{\psi_0},\ket{x})$
hitting-time if the $\ket{x}$-measured walk from this walk and initial state
$\ket{\phi_0}$ has a probability $\geq p$ of stopping at a time $t \leq T$.
\end{definition}

\begin{definition}
We say that the position $x$ is attained in $t$ steps, if $\ket{x}$-measured 
exploration walk has a $(t, 1)$ concurrent
$(\ket{\psi_0}, \ket{x})$ hitting time, i.e. the exploration walk with initial 
state $\ket{\psi_0}$ has a probability of stopping at a time $t<T$ equal to 
$1$.
\end{definition}

With the help of these definitions, one can introduce the concepts of
\emph{visiting strategy} and \emph{attaining strategy}.

\begin{definition}\label{def:visiting-strategy}
If for the given sequence of moves performed by Magnus, there exists $t$ such
that each position on the cycle is visited in $t$ steps, then we call such
sequence of moves a \emph{visiting strategy}.
\end{definition}

\begin{definition}\label{def:attaining-strategy}
If for the given sequence of moves performed by Magnus, each position on the
cycle is attained, then we call such sequence of moves an \emph{attaining
strategy}.
\end{definition}

\section{Application of quantum strategies}\label{sec:application-quantum}
The quantum scheme introduced in the previous section extends the space of
strategies which can be used by both players. As there is a significant
difference in situations where $n=2^k$ and $n=mp$, we will consider these cases
separately.

\subsection{Case $n=2^k$}\label{sec:n=2k}
We start by considering the case $n=2^k$. In this situation we have two possible
alternatives. In the first one Magnus uses the quantum version of the optimal
classical strategy and Derek while Derek performs any 
possible quantum 
moves. In the second scenario both players are able to explore all possible 
quantum moves.

\subsubsection{Quantization of the optimal strategy}
Let us first consider the quantum scheme executed by Magnus with the use of the
classical optimal strategy. As in the classical case Derek is not able to
prevent Magnus from visiting all the nodes, it is natural to ask if he can
achieve any advantage using unitary moves.

If the number of nodes is equal to $2^k$, for some integer $k$, the optimal
strategy for Magnus can be computed at the beginning of the game. This strategy
-- \ie\ a sequence of magnitudes -- is given as (see Lemma~2~in
\cite{nedev08magnus-derek})
\begin{equation}\label{eqn:classical-optimal-2k}
\left(\frac{n}{2^1};\frac{n}{2^2};\frac{n}{2^1}\right);\frac{n}{2^3};\left(\frac{n}{2^1};\frac{n}{2^2};\frac{n}{2^1}\right);\frac{n}{2^4};(:::);\dots ; \frac{n}{2^k};(:::);
\end{equation}
where $(:::)$ denotes the repetition of the moves starting from the beginning of
the sequence until the move preceding the $(:::)$ and excluding it. The first
few sequences resulting from Eq.~(\ref{eqn:classical-optimal-2k}) are presented
in Table~\ref{tab:magnus2-moves-examples}.

\begin{table}[ht]
	\centering
	\begin{tabular}{|l|l|}
    \hline $n$ & optimal sequence of magnitudes \\ \hline
		$2^2$ & $\{2,1,2\}$ \\
		$2^3$ & $\{4,2,4,1,4,2,4\}$ \\
		$2^4$ & $\{8,4,8,2,8,4,8,1,8,4,8,2,8,4,8\}$ \\
        \hline
	\end{tabular}
	\caption{Optimal moves to be performed by Magnus when the number of nodes is
	equal to $2^k$. Magnus is able to visit all $n$ positions in $n-1$ moves by
	using this strategy.}
	\label{tab:magnus2-moves-examples}
\end{table}

By using this strategy in the classical case, Magnus is able to visit all nodes
using $n-1$ moves and Derek is not able to prevent him from doing this.
Moreover, the bound for the number of moves required to visit all the nodes in
the classical case is tight.

Let us now assume that Magnus is using quantum moves constructed for the
classical optimal strategy, but Derek can use arbitrary quantum moves. For
example, if $d=2^3$ Magnus optimal strategy is realized by the following
sequence of unitary gates
\begin{equation}\label{eqn:m-optim-strategy-2k3}
\left\{\left(
\begin{smallmatrix}
 0 & 1 & 0 & 0 \\
 0 & 0 & 1 & 0 \\
 0 & 0 & 0 & 1 \\
 1 & 0 & 0 & 0 \\
\end{smallmatrix}
\right),\left(
\begin{smallmatrix}
 0 & 0 & 1 & 0 \\
 0 & 0 & 0 & 1 \\
 1 & 0 & 0 & 0 \\
 0 & 1 & 0 & 0 \\
\end{smallmatrix}
\right),\left(
\begin{smallmatrix}
 0 & 0 & 1 & 0 \\
 0 & 0 & 0 & 1 \\
 1 & 0 & 0 & 0 \\
 0 & 1 & 0 & 0 \\
\end{smallmatrix}
\right),\left(
\begin{smallmatrix}
 0 & 0 & 0 & 1 \\
 1 & 0 & 0 & 0 \\
 0 & 1 & 0 & 0 \\
 0 & 0 & 1 & 0 \\
\end{smallmatrix}
\right),\left(
\begin{smallmatrix}
 0 & 1 & 0 & 0 \\
 0 & 0 & 1 & 0 \\
 0 & 0 & 0 & 1 \\
 1 & 0 & 0 & 0 \\
\end{smallmatrix}
\right),\left(
\begin{smallmatrix}
 0 & 0 & 1 & 0 \\
 0 & 0 & 0 & 1 \\
 1 & 0 & 0 & 0 \\
 0 & 1 & 0 & 0 \\
\end{smallmatrix}
\right),\left(
\begin{smallmatrix}
 0 & 0 & 1 & 0 \\
 0 & 0 & 0 & 1 \\
 1 & 0 & 0 & 0 \\
 0 & 1 & 0 & 0 \\
\end{smallmatrix}
\right)\right\}.
\end{equation}

\subsubsection{Counterstrategy for Derek}
First of all, as the moves performed by Magnus allow him the sampling of the
space of positions using $r(n)=n-1$ steps, it can be easily seen that Derek is
not able to prevent Magnus from attaining all nodes using $r(n)$ moves.

On the other hand, Derek is able to prevent Magnus from visiting all nodes. He
can achieve this using the strategy given as follows.

\begin{strategy}\label{str:2k-strategy-h-id}
For steps $i=1,\dots,n$ perform the following gate
\begin{equation}\label{eqn:2k-strategy-h-id}
D_i=\left\{
\begin{array}{ll}
H &\mathrm{if}\ i\ \mathrm{is\ odd}\\
\Id &\mathrm{if}\ i\ \mathrm{is\ even}
\end{array},
\right.
\end{equation}
where $H$ denotes the Hadamard gate. 
\end{strategy}

\begin{figure}[ht]
	\centering
    \subfigure[$d=2^3$]{
	    \includegraphics[height=2.1in]{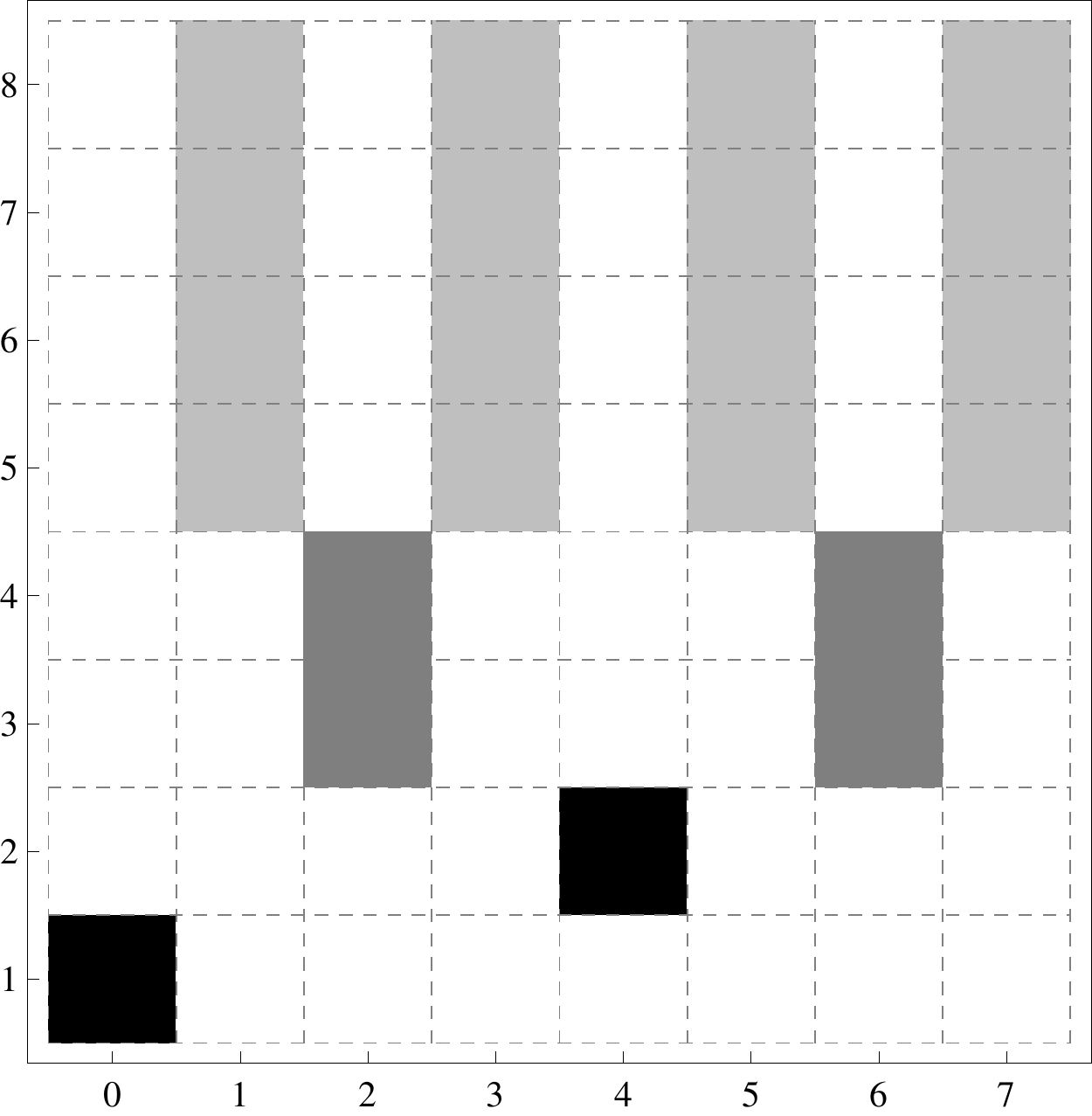}
        \label{fig:qpos-2k3-strategy-h-id}
    }
    \subfigure[$d=2^4$]{
	    \includegraphics[height=2.1in]{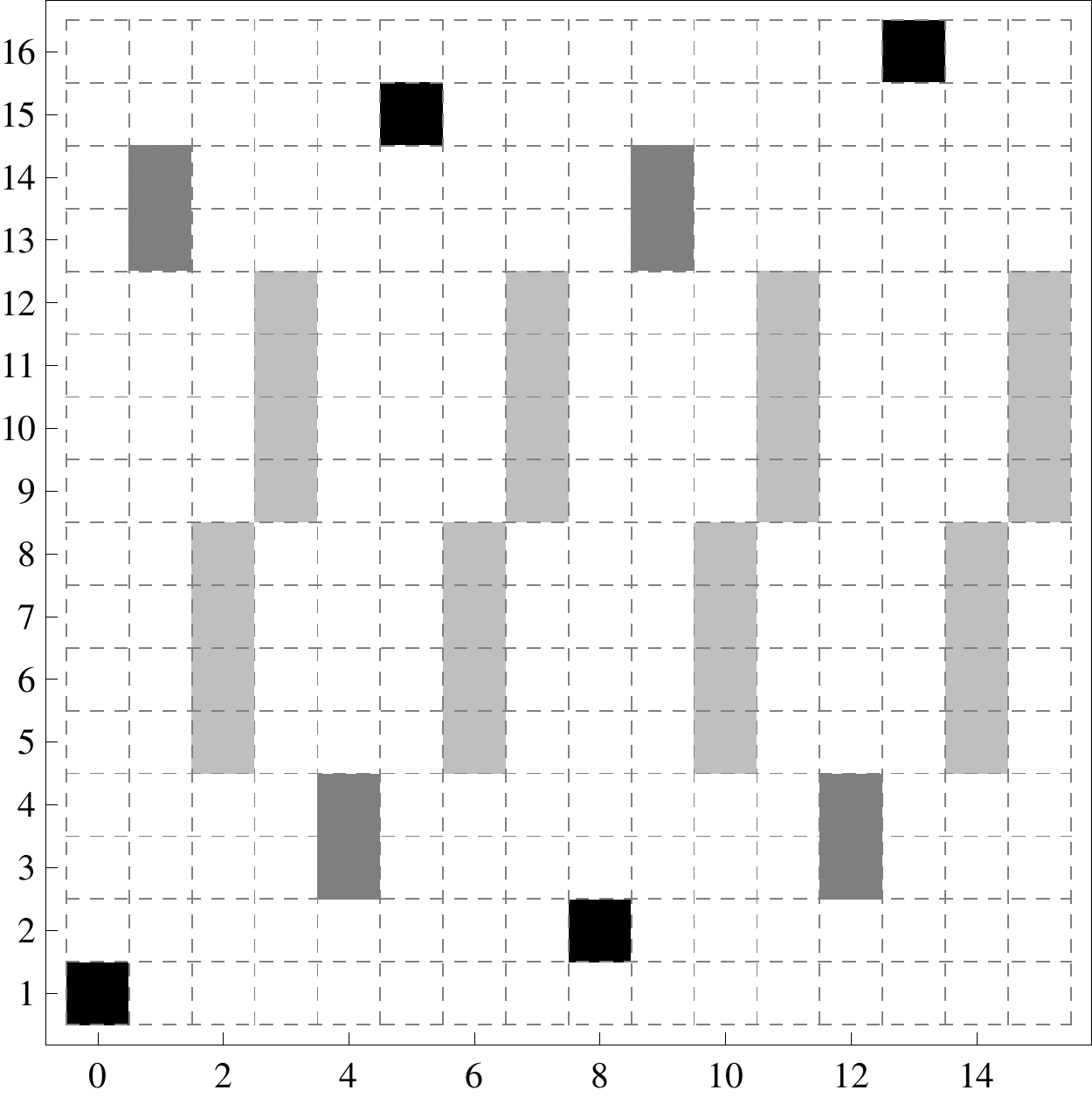}
        \label{fig:qpos-2k4-strategy-h-id}
    }
    \caption{Probability distribution for position in the quantum version of the
    Magnus-Derek game on \subref{fig:qpos-2k3-strategy-h-id} $d=2^3$ and
    \subref{fig:qpos-2k4-strategy-h-id} $d=2^4$ nodes in the case of the Derek
    Strategy \ref{str:2k-strategy-h-id} played against the optimal strategy used
    by Magnus. Gray squares denote attained positions, black squares denote
    visited positions. The higher the grey level, the higher the probability.}
    \label{fig:qpos-2k-strategy-h-id}
\end{figure}

The probabilities of finding a token at each position for the scheme with Magnus
using the optimal strategy and Derek using Strategy \ref{str:2k-strategy-h-id}
is presented in Fig.~\ref{fig:qpos-2k-strategy-h-id}.

Clearly, Magnus is able to attain all the nodes is in $r(d)$ steps.
However, Derek can prevent him from visiting all nodes in $r(d)$ steps. This is
expressed in the following.

\begin{proposition}\label{tw:vs-optimal}
Let us take $d=2^k$. Then, there exists a strategy for Derek preventing Magnus
from visiting all nodes in $r(d)$ steps. Moreover, there is no strategy for
Derek that enables him to prevent Magnus from attaining all nodes in $r(d)$
steps.
\end{proposition}
\proof The first part follows from the construction of the Strategy
\ref{str:2k-strategy-h-id}. In fact, any strategy of this form, not
necessarily using Hadamard gate, will prevent Magnus from visiting all nodes.

The second part follows from the construction of the Magnus' strategy. Let's
assume that there is a strategy that allows Derek to prevent Magnus from
attaining a position $x$ i.e. this position is not attained. Then a
$\ket{x}$-measured walk has no $(n, 1)$ concurrent hitting time. Thus, there is
a non-zero probability that the process will not stop in $n$ steps. 
This means that at each step $t$ there is a non-zero amplitude for some state 
$\ket{p_t} \otimes \ket{m_t} \otimes \ket{d_t}$ with $p_t \ne x$
\ie\ the state will not get measured by effect $\Pi_x$. %
The sequence of directions resulting from the above $d_1, \ldots, d_n$
used by Derek in a classical version of the game would give him a strategy
forbidding a visit in position $x$. It is a contradiction of the properties of
Magnus' classical strategy.

\halmos

The above Proposition can be easily extended as, by using a quantum strategy with
only one Hadamard gate, Derek can prevent Magnus from visiting more than two
nodes.

This result shows that by using quantum moves against the classical strategy,
Derek is not able to exclude additional positions. However, he gains in
comparison to the classical case as he is able to introduce more distraction in
terms of the reliability of the exploration. 

\subsection{Case $n=p m$}\label{seq:n=pm}
In the situation $n=p m$, the quantum strategies used by Derek to distract the
sense of directions can depend on the type of information which is available to
him. Without the possibility to perform position-controlled operations he can
only use classical information about history of choices of Magnus' unitaries
that gives him an estimate of the current state. On the other hand, if he is
able to decide about his move using the current position, the resulting strategy
is more robust.

\subsubsection{Quantum adaptive strategy without position
control}\label{sec:no_position_control}
In the classical case the adaptive strategy allows Derek to use the knowledge
about Magnus' move to choose a step according to the position of a token in 
the moment of the decision. In the quantum case, when a superposition of 
positions is possible and no measurement is allowed, Derek's decision can not 
depend on the position of the token. Instead, Derek can maintain only
information about the history and the current state of the walk in order to 
choose the optimal move. 

In this section we provide such quantum adaptive strategy for Derek under the
principles of the game introduced in Section \ref{sec:quantum-magnus-derek},
\ie\ without using controlled operations, which can be used by Derek to execute
his move. Using the presented strategy Derek can reduce the number of visited
positions to 2 (or even one in the case of odd $n$) at the cost of increasing
the number of attained positions.

The main result of this section can be stated as follows.

\begin{proposition}\label{tw:pm1}
In the case when $n=pqm$%

contains in its 
decomposition two distinct odd 
prime numbers  $p$ and $q$
there exists a strategy for Derek that allows him to assert that:
\begin{enumerate} 
    \item only the starting position (and the symmetric one in the case of even
    $n$) will be visited,
    \item the total number of attained positions during the walk will be at most
    $I=[p(q-1)+1]m=n-(n/q-n/pq)$,
\end{enumerate}
assuming that Magnus uses only permutation operators.
\end{proposition}

One should note that for $n=p^v2^u$ Magnus cannot apply the provided strategy.
Moreover, the strategy could be applied recursively by excluding subsequent
pairs of least odd prime divisors in order to slightly improve this result --
not all multiplications of $pq$ need to be attained and the number of attained
positions would be at most $I'=n-(n/q-n/pq)-(n/pqq'-n/pqp'q')-\dots$, for
$n=pqp'q'\dots2^k$.

In order to prove this, we provide a method for constructing a strategy for
Derek, which allows him to obtain the desired result. We show that the provided
strategy guarantees that the amplitudes of a state, at every step corresponding
to a fixed set of positions, will be equal to zero and, as a consequence, there
is zero probability of measuring any such positions during the walk \ie\ none of
them is attained.

The first requirement for Derek is the choice of the set of \emph{restricted
positions}, \ie\ positions which will be protected from being visited 
or attained by
Magnus. Restricted positions have to be distributed on the cycle in a regular
way. More precisely, we have the following.

\begin{fact}\label{fact:structure}
A set of restricted positions which can be chosen by Derek in order to construct
a strategy in Proposition~\ref{tw:pm1} is a subset of
\[
R_p = p\mathds{Z}_{n/p} = \{ kp : k = 0, .. , \frac{n}{p}-1\},
\]
where $p$ is a divisor of $n$.
\end{fact}

\proof 
To show that the set of restricted position has to be of this form it is
sufficient to prove that the intervals between subsequent restricted positions
have to be equal.

Let us assume that this is not the case and consider three subsequent 
restricted positions. If the distance between two of them is even, Magnus, 
after visiting the position in the middle, would be able to visit one of the 
restricted positions. If both distances are odd, but different, then the sum 
of them is even and by repeating the reasoning track we obtain that Magnus is 
able to visit at least one of the restricted positions. 
\halmos

After choosing the set $R_p$ and for a given position on the cycle, Derek can
choose his move independently from the Magnus' choice. 

When we assign the positions with possible directions according to particular
magnitudes it turns out that some of the positions are not distinguishable from
Derek's point of view. Let us call two positions \emph{symmetric} if their
distance to the nearest restricted position in the direction indicated by the
coin register is identical. This allows us to state the following.

\begin{fact}\label{fact:symmetric}

Considering two symmetric positions the sets of directions that can be 
chosen by Derek in order to avoid visiting restricted positions are identical
for every Magnus' call.
\end{fact}

One can note that the relation of being symmetric is invariant under the action
of the step operator. 

Two facts stated above allow Derek to restrict the choice of moves in such
manner that he is able prevent Magnus from visiting the set of restricted
positions. However, the most important part of Derek's strategy is steering the
state of the system into a superposition of symmetric states. Such a state
guarantees the possibility to perform a strategy in which none of the states
will be visited (only attained).

When such a superposition is achievable from the beginning, Derek achieves the
result similar to the classical case (equal number of restricted positions)
assuming that none of the states is visited. On the other hand, when he needs to
adopt to the standard situation when the starting state is a base state with one
particular position, then the number of states that are attained is greater than
the number of the positions visited in the classical scenario.

\begin{fact}\label{fact:distance}
If a superposition of two symmetric states has been created from a base state,
the beginning position must be equally distant from two closest positions from
every set of the restricted positions.
\end{fact}

\proof 
If this were not the case, the resulting states in a superposition would not be
equally distant from restricted positions and, as the result, not symmetric.
\halmos

The above stated facts allow the formulation of the proof of Proposition~\ref{tw:pm1}.

\proofof{Proposition~\ref{tw:pm1}} 
As a consequence of Fact~\ref{fact:distance}, the starting point has to belong
to every restricted set. Using Fact~\ref{fact:structure}, Magnus can design a
strategy that allows him to visit all positions from an arbitrarily fixed
set $R_p$ (by calling appropriate multiplications of $p$) even when restricted
to the permutation operators. Thus Derek has to choose two prime divisors of $n$
and decide which will be used as the restricted positions set, according to the
Magnus' first move. The optimal choice is to use two smallest factors. In this
case the optimal strategy for Magnus would be to call $p q$ and visit all
positions that are multiplications of $pq$, and then switch to $p$. For this
reason Derek uses the restricted set which is identical as in the case of
restricting $R_q$ excluding all the positions numbered with common
multiplications of $p$ and $q$. From Fact~\ref{fact:symmetric} it follows that
each strategy excluding a given set of positions allows the exclusion of the
same set while operating on a superposition.

Taking into account the above considerations, we define the strategy for Derek,
which fulfills the requirements of Proposition~\ref{tw:pm1}.

\begin{strategy}\label{str:adaptive-no-control}
For any classical strategy used by Magnus, Derek has to perform the
following steps:
\newcounter{step}
\begin{list}{{\rm \bf Step \arabic{step}:}}{\usecounter{step}}
    \item Apply the Hadamard gate.
    \item If Magnus chooses a magnitude equal to $pqk$, for some $k$, apply
    $\Id$ and repeat this step.\\ If Magnus chooses other magnitude go to {\rm
    \bf Step~3}.
    \item If Magnus chooses a multiplication of $p$ (respectively $q$), set the
    restricted positions to be $R=R_q$ (respectively $R=R_p$).
    \item Apply the classical strategy \cite{nedev08magnus-derek}. 
\end{list}
\end{strategy}

Having Strategy \ref{str:adaptive-no-control}, while Magnus applies the 
magnitude equal to $pqk$ and Derek performs Step 2, no positions restricted in 
terms of Prop. \ref{tw:pm1}. will be attained. Starting from the moment that 
Magnus chooses some other magnitude Derek applies unitaries that correspond to 
the classical strategy. This ensures that none of the restricted positions 
will be attained (otherwise Magnus would be able to visit more than $(p-1)n/p$ 
positions in the classical $n=pm$ case, see proof of Prop. 
\ref{tw:vs-optimal}).

\halmos

An example of state evolution in the game executed using
Strategy~\ref{str:adaptive-no-control} is presented in Fig.~\ref{fig:nocontrol}.
The starting position is the only one that is visited. 

\begin{center}
	\includegraphics[height=2.1in]{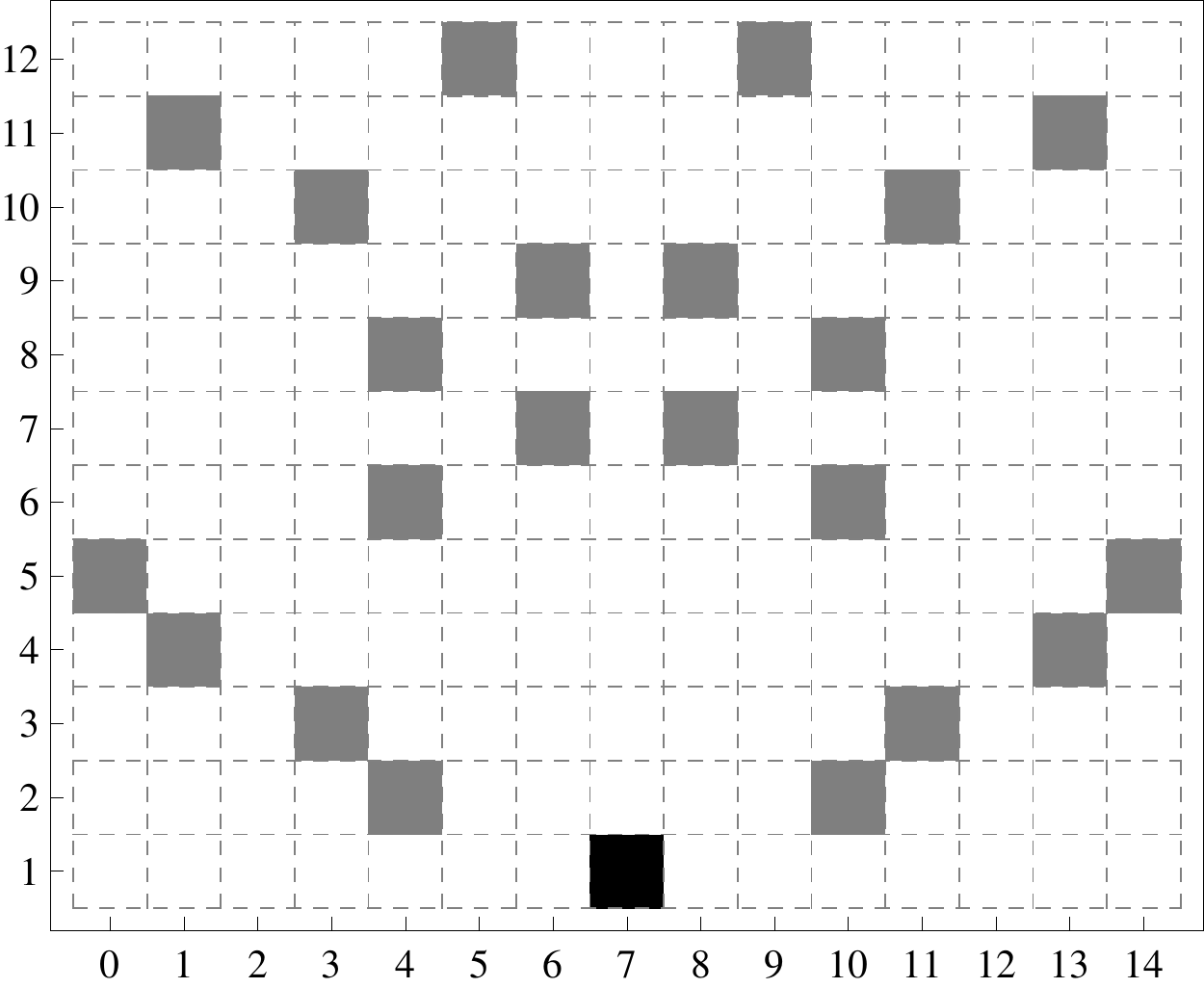}
	\captionof{figure}{An example of the application of the Strategy
	\ref{str:adaptive-no-control} with the initial state $\ket{7}$ on the cycle
	with $n=3\times 5$ nodes. After the first step the amplitudes corresponding
	to the restricted positions (2, 7 and 12) are all equal to $0$.}
	\label{fig:nocontrol}
\end{center}

\subsubsection{Position-controlled adaptive strategy}
\label{sec:position_control}

As it was shown in the previous section a strategy allowed for Derek in the
scenario introduced at the beginning of this paper is not sufficient to maintain
the number of restricted positions characteristic for classical strategy and
limiting Magnus only to attain positions. However, the notion of adaptive
strategies for Derek can be transferred into quantum scenario.

In order to let Derek use position information in his strategy we have to modify
the model introduced in Section~\ref{sec:quantum-magnus-derek}. We do this by
replacing the local operators $\Id\otimes D\otimes \Id$ available to Derek with
the position-controlled operators of the form $\Id\otimes D_p \otimes
\ketbra{p}{p}$. Having such operators at his disposal, Derek is able to apply a
different strategy to each part of the state separately.

\begin{proposition}
Let us consider Magnus-Derek game on $n=pm$, $p>3$, $n\ne 2^k$ positions with
$p$ being the least prime divisor of $n$. When the set of operators available
for Derek includes the operators of the form
\[
\sum_k \Id\otimes D_k \otimes \ketbra{k}{k}
\]
where $k$ is an arbitrary position and $D_k$ is an arbitrary local unitary
operation then the maximum number of attained positions for Magnus is equal to
$n-n/p$ (as in the classical case) and the total number of visited positions is
at most 2 (respectively 1 if $n$ is an odd number).
\end{proposition}

\begin{figure}[ht!]
	\centering
    \subfigure[]{
	    \includegraphics[height=2.1in]{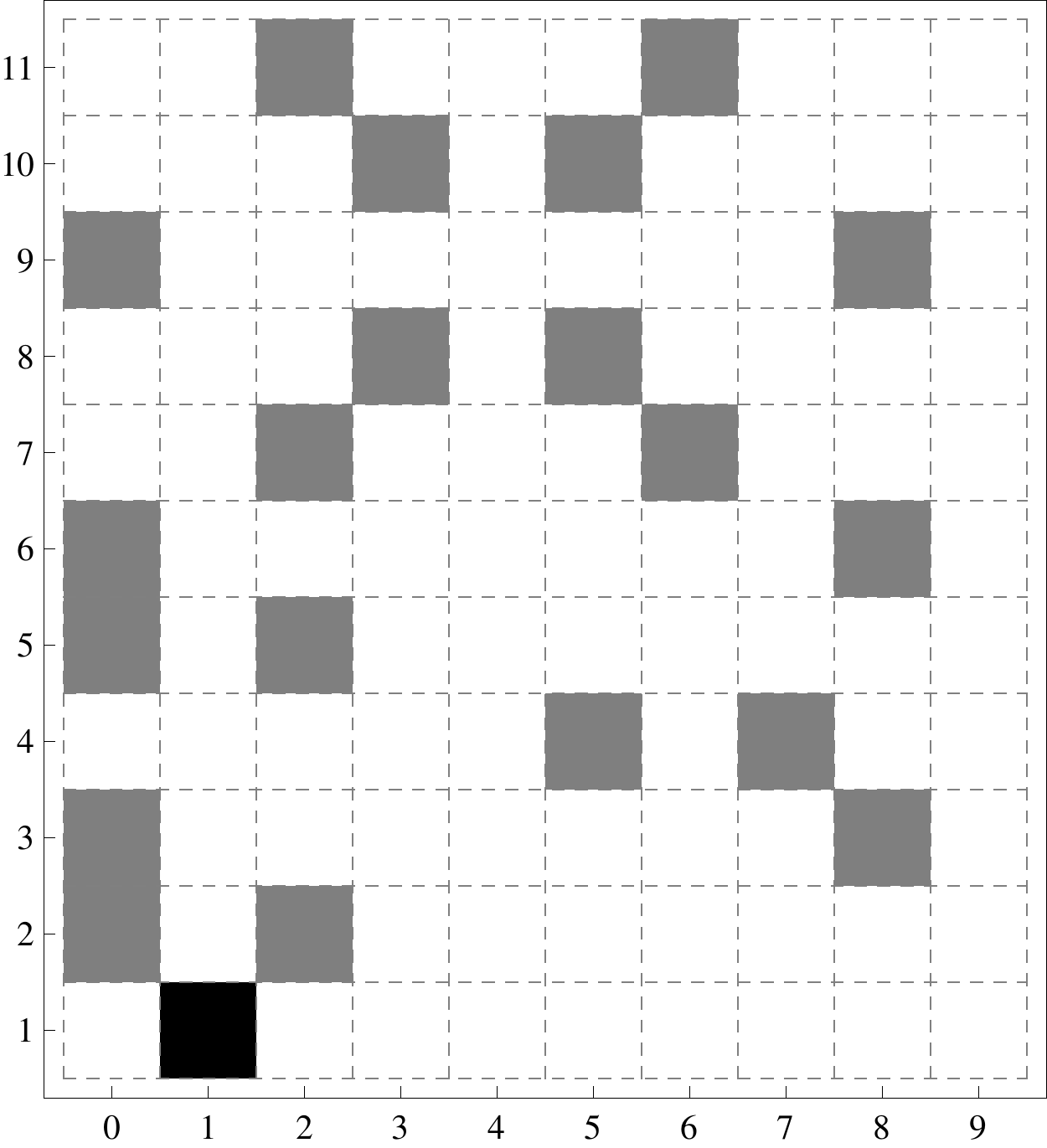}
        \label{fig:controlled-one-hadamard}
    }
    \subfigure[]{
	    \includegraphics[height=2.1in]{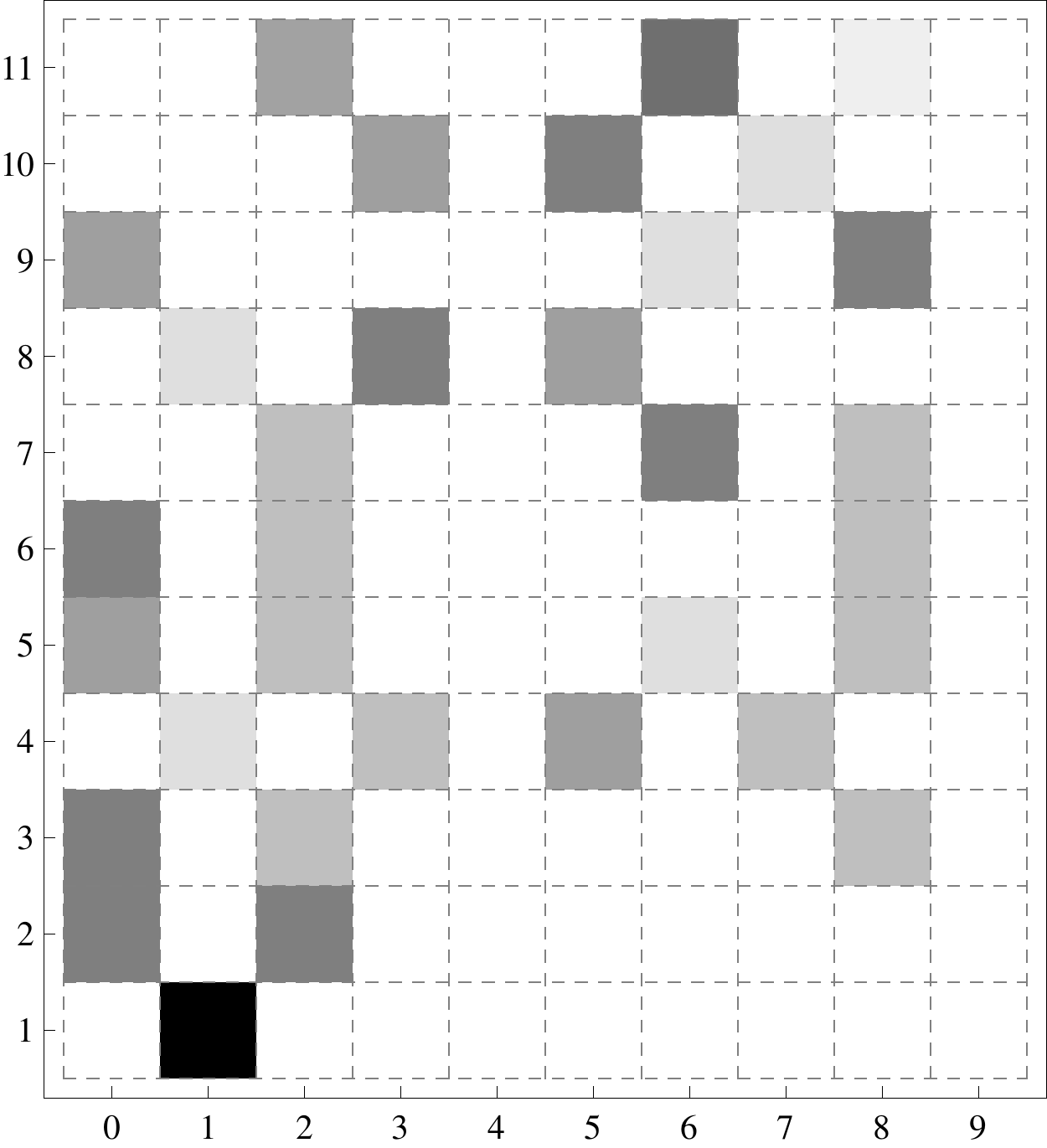}
        \label{fig:controlled-more-hadamard}
    }
    \caption{Examples of application of the Strategy
    \ref{str:adaptive-controled} with the application of the Hadamard gate at
    the first step \subref{fig:controlled-one-hadamard} and Hadamard gates
    applied at steps $1,2,3,4,8,9,10,11$ \subref{fig:controlled-more-hadamard}.
    In both cases the restricted positions are identical.}
    \label{fig:pos_controled}
\end{figure}

\proof
In the simplest case Derek leads to a superposition of two states. In this case
he needs only to ensure that the superposition will not vanish. An example of
such strategy is presented in Fig.~\ref{fig:controlled-one-hadamard}.

\begin{strategy}\label{str:adaptive-controled}
For any Magnus' strategy based on permutation operators, when $n=pm, p>3$ 
and $p$ is a prime number, Derek has to 
perform the following steps:
\begin{list}{{\rm \bf Step \arabic{step}:}}{\usecounter{step}}
    \item Apply the Hadamard gate.
    \item If $n$ is even do nothing as long as Magnus move is equal to $n/2$. If
    $n$ is odd go to {\rm \bf Step~3}.
    \item Find a set of $\frac{n}{p}=m$ equally distant positions that is
    disjoint with already visited positions.
    \item Apply classical strategy to both parts of the state using
    position--controlled operators.
\end{list}
\end{strategy}

The strategy is based on the classical one that is proven to be optimal. 
If there would be a strategy for Magnus that allows him to attain additional 
position, there would be also an analogous classical strategy (see proof of 
Prop. \ref{tw:vs-optimal}).
\halmos

One can also consider a modification of the above strategy with the additional
ability of operating on the superposition of more than two base states. The main
restriction on the strategy executed by Derek is, in this case, the equality of
amplitudes
\begin{equation}
|\braket{m,0,x}{\psi}| = |\braket{m, 1, x}{\psi}|
\vee
|\braket{m,0,x}{\psi}|=0
\vee
|\braket{m,1,x}{\psi}|=0
\label{eq::equal_amplitudes_condition}
\end{equation}
for every position $x$ and current Magnus call $m$. When the condition is
satisfied Derek is able to set an arbitrary direction in every position of the
cycle using the $\Id, H$ and $NOT$ operators. The example is shown in the
Fig.~\ref{fig:pos_controled}. After the second step the state of the token is a
superposition of at least three states. As the consequence, the probabilities
are more distributed over the cycle.

\section{Final remarks}\label{sec:final}
The presented game provides a model for studying the exploration of quantum
networks. The model presented in this paper is based on a quantum walk on a
cycle. Despite its simplicity, the presented model can be used to describe
complex networks and study the behaviour of mobile agents acting in such
network. One should note that in the case of the Magnus-Derek game the main
objective is to optimize the number of nodes visited during the game. The actual
goal of visiting depends on the computation which is required to take place at
the nodes.

We have shown that by extending the space of possible moves, both players can
significantly change the parameters of the exploration. In particular, if Magnus
uses the sequence of moves optimal for the classical case, Derek is able to
prevent him from visiting all nodes. 

We have assumed that in the quantum scenario not only the number of attained
positions is at stake but also the number of positions that are visited by
Magnus. We have considered a modification of a classical strategy that enables
both players to preform their tasks efficiently. This analysis provides an
interesting insight into the difficulty of achieving quantum-oriented goals.

We have also shown that without a proper model of adaptiveness, it is not
possible for Derek to obtain the results analogous to the classical case (the
number of restricted positions is lower or the no-visiting condition is
validated). Performing a strategy optimized in order to reduce the number of
visited slots requires a trade-off with the total number of attained positions.
With additional control resources the total number of attained positions is
maintained if the number of visited positions is strictly limited.

\paragraph{Acknowledgements}
The Authors acknowledge the support by the Polish National Science Centre (NCN)
under the grant number DEC-2011/03/D/ST6/00413. JAM would like to acknowledge
interesting discussions with M.~Mc~Gettrick and C.~R\"over.


\begin{thebibliography}{10}

\bibitem{bernardes00implementation}
M.C. Bernardes and E.~dos Santos~Moreira.
\newblock Implementation of an intrusion detection system based on mobile
  agents.
\newblock In {\em Proceedings of the International Symposium on Software
  Engineering for Parallel and Distributed Systems, 2000}, pages 158--164,
  2000.

\bibitem{chapuran09optical}
T.~E. Chapuran, P.~Toliver, N.~A. Peters, J.~Jackel, M.~S. Goodman, R.~J.
  Runser, S.~R. McNown, N.~Dallmann, R.~J. Hughes, K.~P. McCabe, J.~E.
  Nordholt, C.~G. Peterson, K.~T. Tyagi, L.~Mercer, and H.~Dardy.
\newblock Optical networking for quantum key distribution and quantum
  communications.
\newblock {\em New J. Phys.}, 11:105001, 2009.

\bibitem{chen11more}
L.-J. Chen, J.-J. Lin, M.-Z. Shieh, and S.-C. Tsai.
\newblock More on the {Magnus}-{Derek} game.
\newblock {\em Theor. Comput. Sci.}, 412(4-5):339 -- 344, 2011.

\bibitem{clausen11storage}
Ch. Clausen, I.~Usmani, F.~Bussi\'eres, N.~Sangouard, M.~Afzelius,
  H.~de~Riedmatten, and N.~Gisin.
\newblock Quantum storage of photonic entanglement in a crystal.
\newblock {\em Nature}, pages 508--511, 2011.

\bibitem{feldman10structural}
E.~Feldman, M.~Hillery, H.-W. Lee, D.~Reitzner, H.~Zheng, and V.~Bu\v{z}ek.
\newblock Finding structural anomalies in graphs by means of quantum walks.
\newblock {\em Phys. Rev. A}, 82(4):040301, 2010.

\bibitem{gisin11technology}
N.~Gisin and R.~Thew.
\newblock Quantum communication technology.
\newblock {\em Electron. Lett.}, 46:965--967, 2010.

\bibitem{hurkens08revisited}
C.A.J. Hurkens, R.A. Pendavingh, and G.J. Woeginger.
\newblock The {Magnus}-{Derek} game revisited.
\newblock {\em Inform. Process. Lett.}, 109(1):38 -- 40, 2008.

\bibitem{jansen02intrusion}
W.A. Jansen.
\newblock Intrusion detection with mobile agents.
\newblock {\em Comput. Commun.}, 25(15):1392--1401, 2002.

\bibitem{kempe05discrete}
J.~Kempe.
\newblock Discrete quantum walks hit exponentially faster.
\newblock {\em Probab. Theory Relat. Field}, 133(2):215--235, 2005.
\newblock arXiv:quant-ph/0205083.

\bibitem{kimble08quantum}
H.J. Kimble.
\newblock The quantum internet.
\newblock {\em Nature}, 453(7198):1023--1030, 2008.

\bibitem{lange99seven}
D.B. Lange and M.~Oshima.
\newblock Seven good reasons for mobile agents.
\newblock {\em Commun. ACM}, 42(3):88--89, 1999.

\bibitem{munro11designing}
W.~J. Munro, S.~J. Devitt, and K.~Nemoto.
\newblock Designing quantum repeaters and networks.
\newblock volume 8163, page 816307, 2011.

\bibitem{nedev08magnus-derek}
Z.~Nedev and S.~Muthukrishnan.
\newblock The {Magnus}-{Derek} game.
\newblock {\em Theor. Comput. Sci.}, 393(1-3):124 -- 132, 2008.

\bibitem{papadimitriou}
C.~H. Papadimitriou.
\newblock {\em Computational complexity}.
\newblock Addison-Wesley Publishing Company, 1994.

\bibitem{pembertonross11perfect}
P.J. Pemberton-Ross and A.~Kay.
\newblock Perfect quantum routing in regular spin networks.
\newblock {\em Phys. Rev. Lett.}, 106(2):020503, 2011.

\bibitem{reitzner09searches}
D.~Reitzner, M.~Hillery, E.~Feldman, and V.~Bu\v{z}ek.
\newblock Quantum searches on highly symmetric graphs.
\newblock {\em Phys. Rev. A}, 79(1):012323, 2009.

\bibitem{vanmeter11recursive}
R.~Van~Meter, J.~Touch, and C.~Horsman.
\newblock Recursive quantum repeater networks.
\newblock {\em Progress in Informatics}, (8):65--79, 2011.

\end{thebibliography}
\end{document}